# Determining the Twist Angle of Bilayer Graphene by Machine Learning Analysis of its Raman Spectrum

*Pablo Solís-Fernández, \*,† Hiroki Ago\*,†,‡*

† Global Innovation Center (GIC), Kyushu University, Fukuoka 816-8580, Japan

‡ Interdisciplinary Graduate School of Engineering Science, Kyushu University, Fukuoka 816-8580, Japan

Abstract

With the increasing interest in twisted bilayer graphene (tBLG) of the past years, fast, reliable, and non-destructive methods to precisely determine the twist angle are required. Raman spectroscopy potentially provides such method, given the large amount of information about the state of the graphene that is encoded in its Raman spectrum. However, changes in the Raman spectra induced by the stacking order can be very subtle, thus making the angle identification tedious. In this work, we propose the use of machine learning (ML) analysis techniques for the automated classification of the Raman spectrum of tBLG into a selected range of twist angles. The ML classification proposed here is low computationally demanding, providing fast and accurate results with ~99 % of agreement with the manual labelling of the spectra. The flexibility and non-invasive nature of the Raman measurements, paired with the predictive accuracy of the ML, is expected to facilitate the exploration of the nascent research of tBLG. Moreover, the present work showcases how the currently available open-source tools facilitate the study and integration of ML-based techniques.

Research on twisted bilayer graphene (tBLG) has gained interest in the past few years, sparked in part by the finding of superconductive states at small twist angles.[1] This interest has also expanded to include twisted stacks of other van der Waals (vdW) materials, giving rise to the research field known as twistronics.[2–6] However, determining the twist angle of tBLG or other vdW stacks is not a trivial task. High resolution microscopy techniques like transmission electron microscopy (TEM)[7–12] or scanning probe microscopies (SPM)[13–16] provide the most accurate angle determinations, with precisions below ~0.01°. The downside is that measurements are time-consuming and require either a free-standing sample or supported on a conductive substrate. Moreover, they provide very local information on sub-micron sized areas, whereas the twist angle can vary considerably within a few µm.[17,18] These techniques are thus not suitable for practical applications, which require large-area characterizations on arbitrary substrates and in relatively short times. Similar accuracies can be obtained from transport measurements of devices under magnetic fields at low temperatures,[1] but the complexity of the measurements and their limitation to small areas make them unsuitable in practical applications. Low-energy electron diffraction (LEED) can provide information about the number of layers and their stacking orientations with sub-degree accuracy for areas that can be larger than those obtained from TEM and SPM, although generally requires a conductive substrate and high vacuum conditions.[2,17–19] In the case of isolated tBLG grains with hexagonal shapes it is also possible to determine the twist angle by optical microscopy.[20] This technique is simple and provides an accuracy of ~1°, but it cannot be applied to the case of tBLG with high coverages or with non-hexagonal shapes, as generally occurs for graphene grown by chemical vapor deposition (CVD).[18]

Raman spectroscopy is a non-invasive analysis technique that provides flexibility about the kind of substate and environment in which the measurement is done, along with the possibility to examine relatively large areas in reasonably short periods of time. It has been widely used in the characterization of graphene, providing extensive information about its characteristics, quality, and electronic state.[21] In particular, the Raman spectrum of graphene depends on the number of layers and on their relative orientation.[7–9,11,20–25] The case of Bernal and rhombohedral few-layer graphene (FLG) has been extensively studied owing to their high occurrence in natural graphite. For it, both the shape of the Raman 2D band and the breathing modes in the spectral region at 1650 – 1800 cm-1 are good indicators of the stacking order.[23,24,26] For the case of tBLG it also results possible to determine the twist angle by Raman spectroscopy, providing even sub-degree accuracies for certain angle ranges.[27] In general, determining the twist angle requires the simultaneous comparison of several features of the Raman spectrum. However, the increased complexity of the spectra can greatly difficult this task.[7,8,10,11,17,25] Although information on the twist angle of tBLG is encoded in the Raman spectrum, the variations for different angles can be very subtle, often involving small changes in the positions, widths, and/or intensity ratios of the different peaks. These are frequently imperceptible at first glance and might easily be overlooked, requiring a careful inspection of the spectrum.[11,20,27] Observing some of them rely on the obtention of high quality spectra with large signal-to-noise ratios, as it is case for the subtle changes in the spectral region between the G and 2D bands or at low Raman wavelengths.[20,25,27,28] Moreover, the Raman spectrum can be significantly altered by the electronic, chemical and structural state of the graphene, such as due to the presence of defects, strain or doping.[21,29–33] This further make it difficult to determine the twist angle, as each angle is represented by a broad dispersion of spectra instead of by a single ideal spectrum. All these make the manual determination of the stacking order a time consuming and not practical task. As the twist angle of current CVD-grown tBLG can greatly vary point to point,[17,18] it is necessary to introduce automated Raman-based analysis methods for the fast and reliable determination of the stacking order for large areas.

Machine learning (ML) comprises a series of techniques that rely on statistics to categorize new data based on the training of a model (supervised ML), or to find patterns in uncategorized data (unsupervised ML).[34] ML-based methods are being actively introduced in different aspects of the research and handling of 2D materials.[35–38] Recently, ML were proved to be effective in determining the twist angle of simulated portions of the Raman spectrum mainly around the G band.[39] However, real Raman spectra can differ significantly from the simulated ones, with the positions, widths, and relative intensities of the most characteristic Raman peaks being heavily affected by things like strain and doping.[21] ML has also been used to identify limited twist angles of BLG made by artificial stacks of SLG.[40] Clustering, an unsupervised ML paradigm, has also been used to assemble uncategorized Raman data into groups roughly corresponding to different kinds of BLG, but with no sensitivity towards the twist angle.[41]

In this work, we propose an easy, fast, and low computationally demanding ML-based method to determine the stacking order of tBLG from its Raman spectrum. As shown in Figure 1, this method involves extracting selected features of the Raman spectra of tBLG that can provide enough information to train a ML model to infer the twist angle within some predefined ranges.

The accuracies of the ML predictions can exceed a 99 % of agreement with the manual labelling of the spectra. Combining this predictive accuracy with the flexibility and non-invasive nature of the Raman measurements is expected to facilitate and accelerate the nascent research of twisted vdW stacks. The present work also introduces an example of how currently existing available open-source tools allow an easy and effective integration of ML-based techniques with regular research methods, potentially increasing the efficiency of the research.[42]

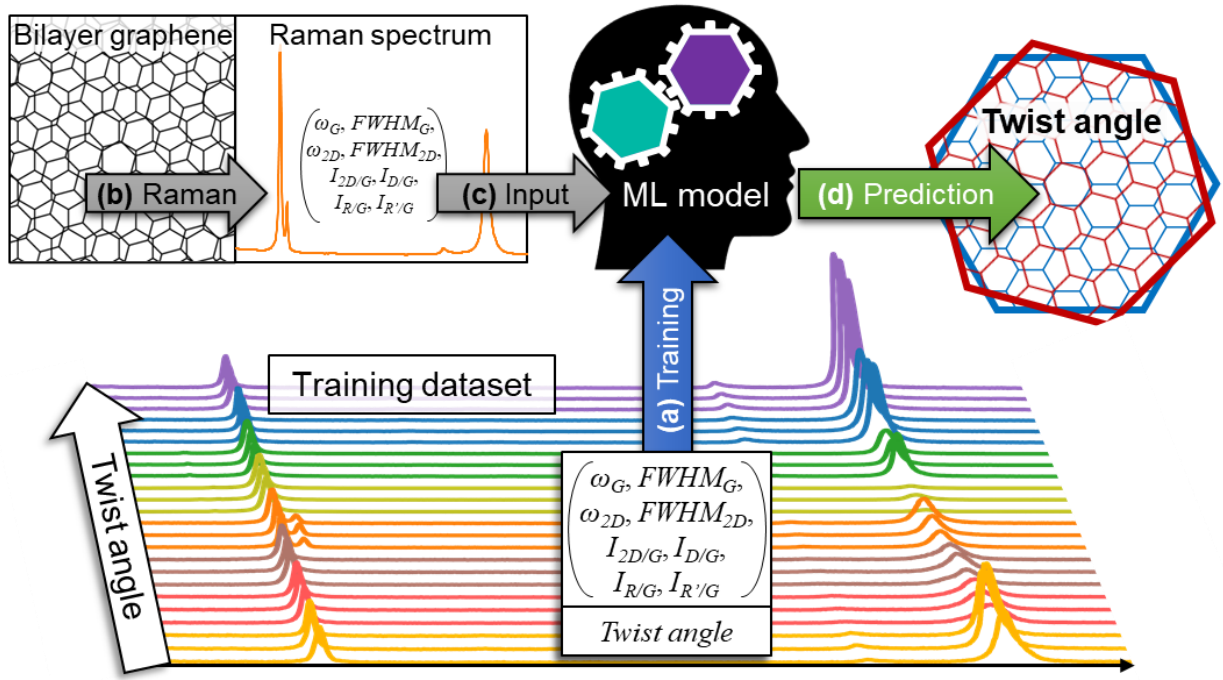

**Figure 1**. Schematic of the process involved in the ML determination of the twist angle of BLG. (a) A ML model is trained using data extracted from a set of Raman spectra for which the twist angle was previously determined (training dataset). (b) Raman features are collected from a different tBLG sample for which the twist angle have to be determined. (c, d) The trained model can then infer the twist angle of other BLG samples from their Raman signature.

**Results and Discussion**

Determining the twist angle of tBLG by Raman. Several of the characteristics of a given graphene sample can be obtained by the inspection of its Raman spectrum, including the strain, doping, presence of defects, or thickness.[21,31] The twist angle of tBLG can also be roughly estimated from the Raman characteristics.[11,20] Figure 2 shows the average spectra for CVD-grown SLG, AB-stacked BLG (BLG-AB), and for tBLG with different twist angle ranges, all of them after being transferred to $SiO_2$ substrates. The averages were obtained from ~6000 Raman spectra shown in Fig. S1, which were used without any further preprocessing, such as noise reduction or background subtraction. These spectra were collected from graphene grown in different CVD batches, to obtain the wider possible range of angles, and to counteract small sample to sample differences and unintentional variations of strain or doping levels.[18] Careful inspection of each individual Raman spectrum allowed us to assign them to one of eight different classes shown in Fig. 2. This classification was done by comparing the specific characteristics of each spectrum with spectra reported in the literature and precisely labelled by alternative experimental techniques.[11,20] This manual process to classify the Raman spectra is laborious and prone to errors, and given that the twist angle can greatly vary over the BLG surface,[18] it is difficult to scale to large CVD-grown BLG areas.

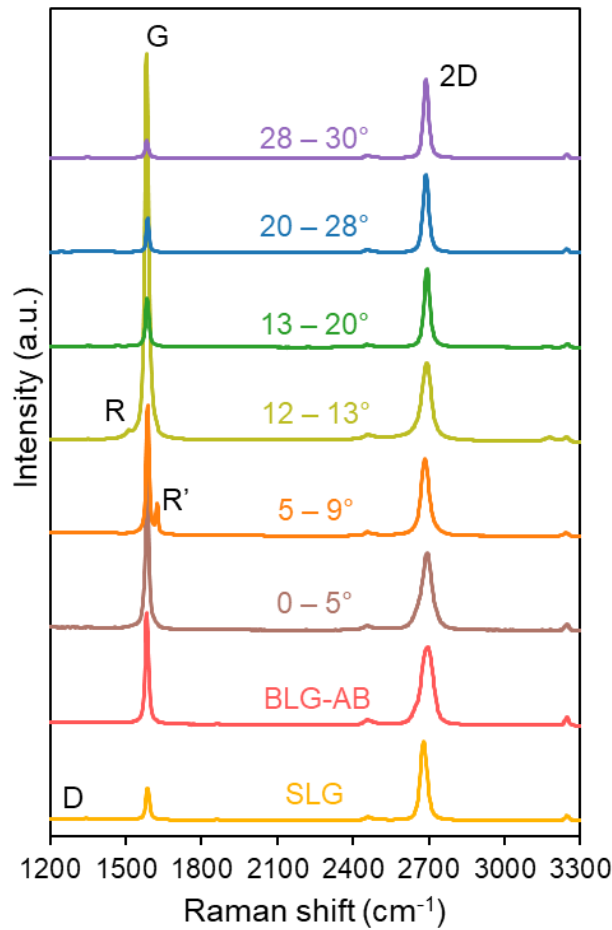

**Figure 2**. Average spectra of SLG, BLG-AB and tBLG with different twist angles, measured using a 532 nm excitation. Each spectrum corresponds to one of the classes included in the training dataset for the ML models. The twist angles were determined by comparison with existing literature. The spectra are normalized to the 2D band intensity and vertically shifted for clarity, and the most characteristic Raman bands are indicated with labels.

**Table I**. Average values of the Raman features for each of the twist angle ranges used to train the ML models. The last row shows the mutual information between each of the features and the classes of the training dataset.

| Twist angle | $\overline{\omega}_G (cm^{-1})$ | $\overline{FWHM}_G (cm^{-1})$ | $\overline{\omega}_{2D} (cm^{-1})$ | $\overline{FWHM}_{2D} (cm^{-1})$ | $\overline{I}_{2D/G}$ | $\overline{I}_{D/G}$ | $\overline{I}_{R/G}$ | $\overline{I}_{R'/G}$ |
|---|---|---|---|---|---|---|---|---|
| SLG | 1587.02 ±3.17 | 14.82 ±1.45 | 2680.27 ±3.31 | 30.15 ±2.2 | 2.18 ±0.5 | 0.07 ±0.04 | | |
| BLG AB | 1583.35 ±2.08 | 15.17 ±1.15 | 2694.60 ±2.5 | 53.55 ±1.25 | 0.66 ±0.1 | 0.04 ±0.01 | | |
| 0–5° | 1584.24 ±0.65 | 15.34 ±0.79 | 2693.05 ±1.21 | 45.03 ±2.21 | 0.46 ±0.03 | 0.03 ±0.003 | | |
| 5–9° | 1586.77 ±2.05 | 13.31 ±1.69 | 2685.43 ±2.39 | 35.55 ±2.3 | 0.57 ±0.14 | 0.04 ±0.01 | | 0.25 ±0.12 |
| 12–13° | 1582.91 ±3.23 | 15.89 ±2.29 | 2691.08 ±5.66 | 37.65 ±4.92 | 0.19 ±0.12 | 0.01 ±0.004 | 0.04 ±0.01 | |
| 13–20° | 1584.98 ±1.83 | 15.41 ±1.73 | 2692.69 ±2.01 | 25.66 ±2.42 | 1.55 ±0.48 | 0.04 ±0.01 | 0.05 ±0.01 | |
| 20–28° | 1587.04 ±1.2 | 12.03 ±1.45 | 2688.75 ±2.43 | 25.20 ±1.88 | 2.26 ±0.36 | 0.07 ±0.02 | | |
| 28–30° | 1583.44 ±1.34 | 15.39 ±1.28 | 2688.84 ±2.61 | 23.03 ±2.01 | 4.47 ±0.49 | 0.07 ±0.03 | | |
| MI | 0.390 | 0.295 | 0.769 | 1.220 | 1.266 | 0.613 | 0.182 | 0.284 |

Once that the spectra are classified, we extracted a few measurable features from each spectrum (Fig. 1).[40,41] The selected features include the shift ($\omega$) and full-width at half maximum (*FWHM*) of each of the main graphene Raman bands (G and 2D), the relative intensities of the 2D, D, R and R' bands respect to the G band ($I_{2D/G}$, $I_{D/G}$, $I_{R/G}$ and $I_{R'/G}$). The relative band intensities were used instead of the actual intensities, as they are less prone to variations from different measurements. The ratios $I_{R/G}$ and $I_{R'/G}$ were only collected for spectra with noticeable R and R' bands, while for the rest they are considered to be 0. All the features are represented in Figure S2, with their average values shown in Table I for the different assigned twist angle ranges. The last row of Table I corresponds to the values for the mutual information between each feature and the assigned classes for the twist angle, which is a measure of the shared information between the corresponding feature and the twist angle.[43] Hence, the features with the largest values are expected to be the most relevant to determine the twist angle. Taken separately, none of these features allow to unambiguously determine the twist angle. This even happens for the two features with the largest mutual information, for which Figure 3a,b show the corresponding values of all the spectra separated by classes. The intensity ratio $I_{2D/G}$ is not sensitive to angles between 0° (BLG-AB) and 9° (Fig. 3a), with data points for these classes overlapping in the range $0.5 \lesssim I_{2D/G}$

⪅ 1. On the other hand, *FWHM$_{2D}$* fails in discriminating twist angles between 13° and 30°, for which most of the data points lie in the range 20 ⪅ *FWHM$_{2D}$* ⪅ 30 (Fig. 3b).

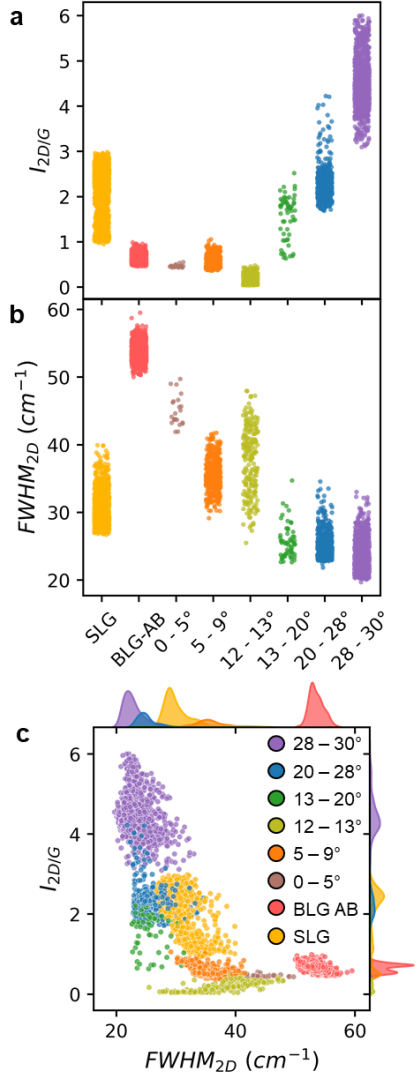

**Figure 3**. (a, b) Distribution of *I$_{2D/G}$* (a) and *FWHM$_{2D}$* (b) for the different classes. (c) Scatter plot of the *FWHM$_{2D}$ vs* the intensity ratio *I$_{2D/G}$* for each of the classes. The probability distributions are included at the top (*FWHM$_{2D}$*) and right (*I$_{2D/G}$*) edges of (c).

Improving the sensitivity consequently requires the simultaneous use of features that can provide complementary information. This can be seen when plotting pairs of features (Figure S3). The specific case of *I$_{2D/G}$* and *FWHM$_{2D}$* (Fig. 3c) shows how each of the classes forms a cluster. The presence of these clusters provides a faster way to label new experimental data as compared with the manual inspection of the whole spectrum. But although the differentiation of the classes is now clearer than in Fig. 3a,b, as shown in Figure S4 some of these clusters still overlap, preventing an unequivocal determination of the twist angle by simply using two features. The overlap of the clusters is reduced by further increasing the number of features ($n$), as they will now be embedded in a $n$-dimensional hyperspace. By combining all the available features shown in Fig. S2 ($n = 8$), it is thus possible to increase the accuracy in the determination of the twist angle. To confirm this, we tried to determine the classes of the graphene region shown in Figure 4a. The Raman maps for $I_{2D/G}$ and $FWHM_{2D}$ are shown in Fig. 4b,c, with the maps for the whole set of features in Figure

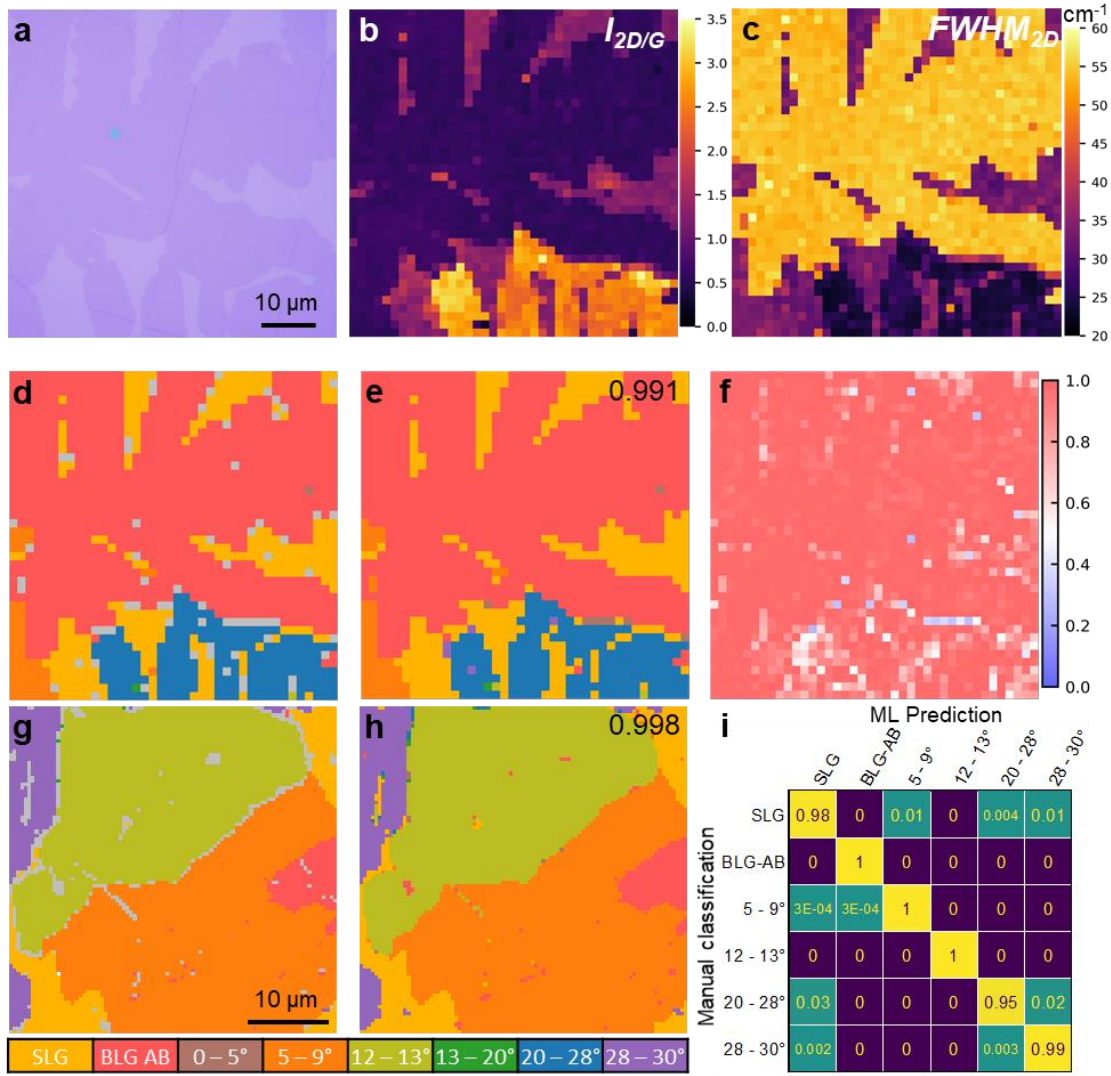

**Figure 4**. (a) Optical image of CVD-grown graphene consisting of SLG (light) and BLG (dark) areas. (b, c) Raman mappings of $I_{2D/G}$ (b) and $FWHM_{2D}$ (c) for the area in (a). (d, e) Mapping of the twist angle determined by manually inspecting the characteristics of the Raman spectra (d) and by a trained random forest ML model (e). The predictions in (e) coincide in ~99.1 % (value in the top right corner) with the manual determination in (d). This value excludes the points in (d) that could not be manually labelled (gray). (f) Confidence score of the random forest model to each of the predictions shown in (e). (g, h) Manual classification (g) and ML prediction (h) of a different area (accuracy ~99.8 %). (i) Confusion matrix for the ML predictions in (e) and (h), normalized by rows.

S5. The contrast in the Raman mappings indicates the existence of different graphene regions. Combining the data from all these maps allows to manually assign a class to each of the spectra, as shown in the Fig. 4d. Most of the spectra of the mapping could be classified this way, except for the few points marked in grey that are usually found in limits between two different regions. The manual assignment was also performed on other areas comprising each of the classes (Figure S6). Following this method, it was thus possible to distinguish not only between SLG and BLG, but also different twist angles for the latter. But this approach still requires a considerable manual input and a careful inspection of the Raman data in an $n$-dimensional hyperspace. Moreover, it can also hinder the identification of finer details that might potentially go unnoticed, thus preventing to reach a higher precision in the determination of the twist angle. An alternative is to decrease the dimensionality of the feature hyperspace.[44] When performed carefully, the lower dimensional space can retain most of the original information of the original while making the analysis simpler (Figure S7).

**Training a ML model to determine the stacking order.** To increase the processing speed and precision of the labeling of the Raman spectra we applied ML methods. The $n$-dimensional vector data conformed by the chosen features of a Raman spectrum can be easily interpreted and processed by relatively simple ML models. As we are suing supervised learning, the first step is training the chosen model using Raman data that has already been labelled (Fig. 1a). The models used here are multiclass classifiers for which the training data is categorized into a finite set of classes, corresponding to the previously shown SLG, BLG-AB, and tBLG with several twist angle ranges (Fig. 2). The dataset used to train the model was the same used in the manual classifications, by assigning a class to the spectra in Figure S1 and extracting the relevant Raman features (Fig. S2 and Fig. S3). The training ultimately consists of feeding the ML model with both the values of the Raman features and the manual classification of each spectrum. Each model has a different set of internal variables, known as hyperparameters, that have a large impact on the performance of the model but cannot be derived during the training process. As specified in the Methods section, these hyperparameters were optimized for each different model by cross-validation.

Once that it has been trained, the model is ready to predict the twist angle of tBLG from its Raman spectrum (Fig. 1b-d). Owing to the inclusion of a class for SLG in the training dataset, the model is also able to discriminate between BLG and SLG. It is important to note that all the ML predictions done in this work are for data that was not present in the training set, being collected from different CVD batches. This ensures that the model can generalize to new unseen data, and that the high predicting accuracies are not due to problems like overfitting of the training data.[45] To carry out the prediction, the $n$-dimensional vector containing the Raman features of data unseen by the trained model is used as input. In the case of the area shown in Fig. 4a, the trained model was supplied with the maps containing the Raman features for each point (Fig. 4b,c and Fig. S5). The prediction using a random forest model is presented in Fig. 4e, matching ~99.1 % with the values obtained by manual labelling (Fig. 4d). It was not possible to find a single area of graphene containing all the different classes that can be predicted. Fig. 4g,h show the manual classification and the model prediction of another area. Both areas together include all the classes but $0 - 5°$ and $13 - 20°$, which are the least frequent twist angles. Finally, all the classes are represented when including the areas shown in Fig. S6, each collected from a different sample. The average accuracy of the model predictions for the areas in Fig. 4e,h and Fig. S6 is ~99.0 %, obtained as a weighted average to account for differences in the size of the areas. Fig. 4i show the confusion matrix for the predictions in Fig. 4e,h, while the confusion matrix in Figure S8 summarizes the predictions from all the measured areas. The confusion matrix evidences that all the classes achieve good accuracy ratios, with most of the mistakes make by the ML model being due to confusion between the SLG and tBLG with twist angles of $20 - 30°$. The model used in these calculations is a random forest classifier, as is a commonly employed model that performs well on a wide range of problems, and it is relatively easy to understand.[46] The performance of other ML algorithms was also studied, with the results of the predictions being shown in Figure S9. After adjusting their corresponding hyperparameters, all the selected models clearly outperform the results obtained from the dummy models that were also included in Fig. S9.

With some exceptions, during the prediction the ML models assign a probability for the input data to belong to each of the classes. The class predicted by the model is that with the highest probability, known as the confidence score and which can be interpreted as a confidence of the model on the prediction. Fig. 4f shows the confidence score of the random forest model to the predictions in Fig. 4e. The high value of the average confidence score, ~0.95, is consistent with the high accuracy of the prediction. The areas with the lower probabilities mostly coincide with areas in which the manual labelling was not possible, such as in the boundaries between different stacking regions, and with areas in which the predictions are more likely to disagree with the manual labelling.

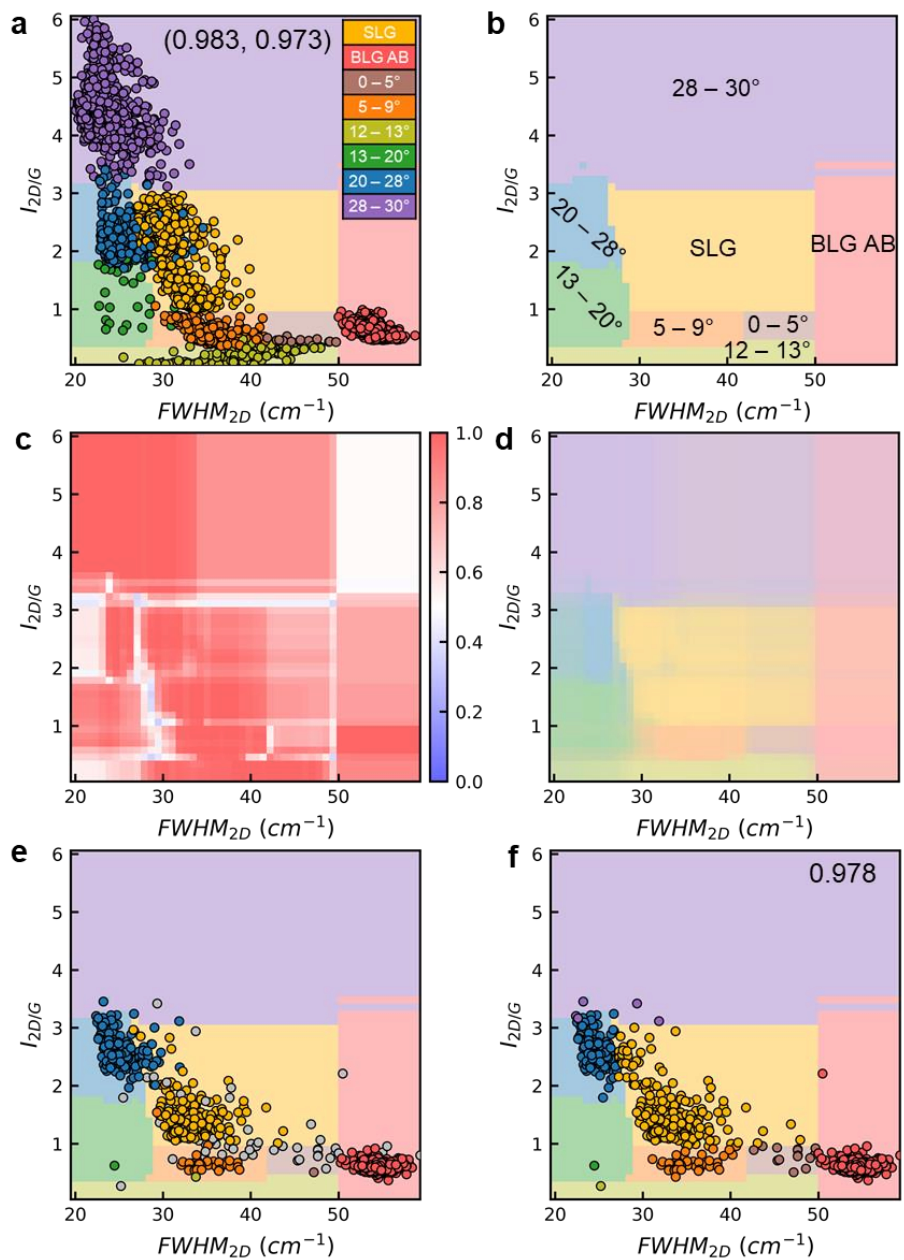

**Figure 5**. (a) Scatter plot of the $FWHM_{2D}$ vs $I_{2D/G}$ used for training a random forest classifier, with the numbers in parenthesis represent the accuracy of the model on the train and test sets, respectively. The colored background shows the corresponding decision boundary of the model. (b) Decision boundary for the random forest classifier trained with the data shown in (a). (c) Confidence score of the model for each point of the decision boundary. (d) Decision boundary with the colors determined according to the weighted average of the probabilities assigned to each of the classes. (e, f) Decision boundary including scatter plots for the maps in Figures 4b,c with the classes manually assigned (e), and predicted by the model (f). The number in the top right corner of (f) represents the accuracy of the model with respect to the manual labelling, excluding the points in (e) that could not be manually labeled (gray colored).

To better understand the training and prediction processes, a random forest model was trained using only the pair of features with the largest values of mutual information, i.e., $I_{2D/G}$ and $FWHM_{2D}$, as shown in Figure 5a. This decreases the accuracy of the predictions but allows to visualize the decision boundary of the model, which divides the feature space ($FWHM_{2D}$, $I_{2D/G}$) into regions for which different classes are predicted. The resulting decision boundary is displayed as the colored background of Fig. 5a,b, obtained by plotting the class with the largest probability estimated by the model on each point of the feature space. The fact that the decision boundary appears simple indicate that the model is not overfitting the training set.[45] This is in accordance with the high accuracy value obtained on the hold-out test set, and on the predictions of Fig. 4e,h and Fig. S6. For comparison, in Figure S10 there are examples of decision boundaries for overfitted and underfitted support vector machine (SVM) models. The overfitted model shows a complex decision boundary that closely follows the data from the training set. Consequently, it has a perfect accuracy on the training set, but the much lower value obtained in the test set indicates that it does not generalize well to predict new results. Conversely, the underfitted model shows extremely simple boundaries, with similarly low accuracy values for both the training and the test sets. These results stress the importance of choosing suitable hyperparameters to obtain reliable predictions from the models.

The value of the confidence score for each point of the decision boundary is shown in Fig. 5c. Regions with low probability values can be found in areas void of training data, as in the top right region. Nevertheless, this does not affect the prediction accuracy of the model, and only reflects the uncertainty in areas of the feature space for which experimental data does not physically exist (e.g., graphene does not concurrently show broad and intense 2D bands). Especially problematic for the model predictions are the low probability values in boundaries in which training data for different classes are close to each other. This highlights the need for properly trained models that can maintain high accuracies on such regions. The uncertainty of the model is also represented in Fig. 5d, for which the color of each point is a weighted linear combination of the color of each class and the probability assigned to it. In contrast with the sharp boundaries of the decision boundary (Fig. 5b), the boundaries between different classes are now blurred, reflecting the low confidence of the model for the predictions in these regions.

Figures 5e,f show how the model trained only with $I_{2D/G}$ and $FWHM_{2D}$ classifies the Raman mapping from Fig. 4a. In Fig. 5e] the classes are manually assigned (Fig. 4d), while Fig. 5e shows the result of the ML prediction (Figure S11). It is evident that the class of some of the dots in Fig. 5e do not correspond with the region of the decision boundary on which they stay. This happens for some of the dots assigned a 20 – 28° twist (blue dots) but located at the lower part of the 28 – 30° region (purple area) and at the top left part of the SLG region (yellow area). Given the way that the decision boundary plot is constructed, the prediction from the model trivially follows the regions delimited by the boundaries, and hence those dots are being mislabeled by the model and decrease its accuracy. This mislabeling mainly occurs close to the boundaries delimiting the regions, which were previously mentioned to have low confidence scores (Fig. 5c). Another thing to note is that model can assign a class for the dots that could not be manually classified (gray in Fig. 5e). This is especially evident in Fig. 5e for the unlabeled dots found in the 0 – 5° region (brown area). Although it is not possible to ascertain the accuracy of the model in these cases, the outcome seems plausible given that these dots are close to areas of BLG-AB in Fig. S11c. Ultimately, the confidence score of each prediction can be used as a threshold value to discard ML predictions with low scores.

**Limitations of the ML models.** It is important to understand the limitations of the proposed ML approach and the factors that influence the accuracy of the predictions. One of the most important factors in the ML analysis is the quality of the training data. Here we present the impact of the chosen Raman features and of the size of the training set. As previously mentioned, not all the features have the same relevancy on the ML accuracy. To see this in a practical case, we compare the results of a manual labeling (Figure 6a]) with the predictions of models trained with only two features. When choosing two features with large values of mutual information, such as $I_{2D/G}$ and $FWHM_{2D}$, the model retains a relatively good accuracy of ~97.8 % (Fig. 6b and Fig. S11), only slightly lower than the value obtained when using the complete set of features (~99.1 %). In contrast, selecting less relevant features, such as $\omega_G$ and $I_{D/G}$, decreases the accuracy to ~77.9 % (Fig. 6c and Fig, S11). This highlights the importance of selecting features that are adequate to the problem in order to obtain the best possible outcome.

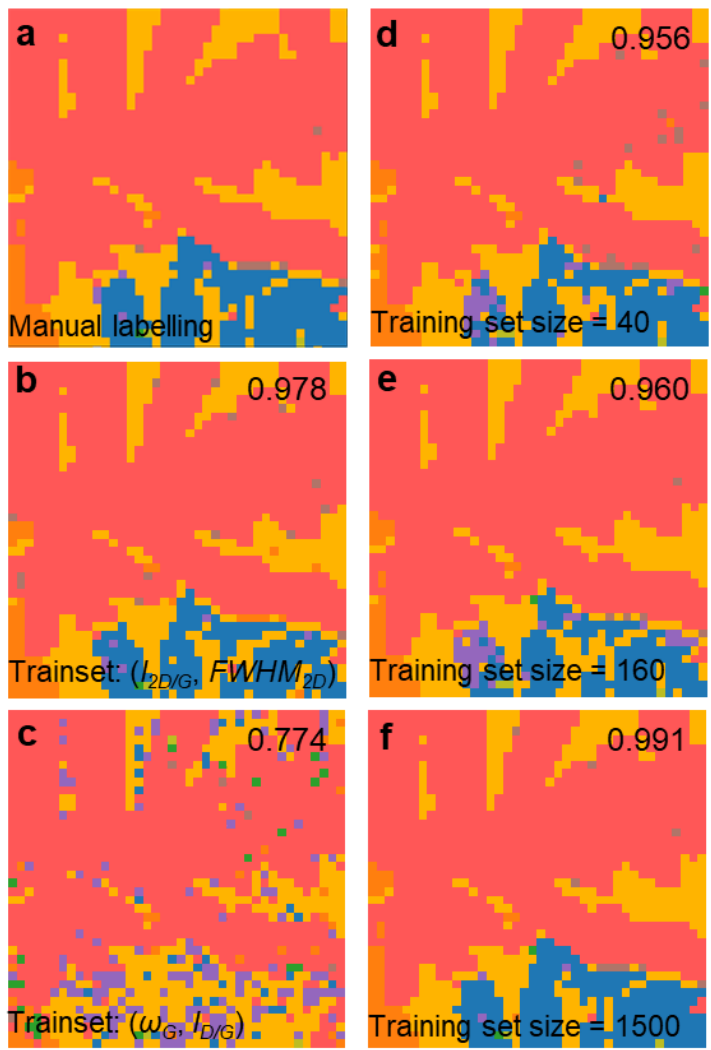

**Figure 6**. Effect of the features and the size of the training dataset on the predictions of a random forest model. (a) Manual labeling. (b-c) Predictions for a model trained using the features ($FWHM_{2D}$, $I_{2D/G}$) (b), and ($\omega_G$, $I_{D/G}$) (c). (d-f) Predictions for models trained with dataset sizes of 40 (c), 160 (e) and 1500 (f) instances. The numbers in the top right corner represent the accuracy of the model in each case for this specific area.

The ideal size of the training dataset depends on the specific problem and of the model used, with the most complex cases requiring datasets with several thousands of instances. Our training set was obtained from ~6000 spectra, with the accuracy being expected to decrease for smaller sets. This is evidenced in the evolution of the predictions in Fig. 6d-f for increased sizes of the training set. Figure S12a shows the accuracy of a random forest model for different sizes of the training set. The accuracy was obtained for the same mappings shown in Fig. 4e and Fig. S6. Even for the smallest sets tested, with 40 to 160 instances, relatively high accuracies of ~91 % were attained. This shows that modest set sizes can still provide reasonably good predictions of the stacking order. As shown by the error mappings in Fig. S12, most of the errors of these smaller sets arise from the confusion between classes with similar spectra, e.g., between BLG-AB and 0 – 5°, or between 20 – 28° and 28 – 30°. Thus, for small sets the accuracy can be potentially increased by focusing in obtaining additional training instances of such classes. A steep rise of the accuracy is observed when the training set is increased, reaching a steady value of ~91 % for sizes over ~1500. This indicates that the best strategy to further increase the accuracy for these set sizes can be to improve the quality of the dataset rather than continue increasing its size.

The high sensitivity of the Raman spectra to the state of graphene can also limit the accuracy of the ML determination of the twist angle. This is due to the spectra included in the training dataset are acquired under certain specific conditions, so that the model might be unable to properly classify spectra acquired under different conditions. It is thus important to stablish the limits of the training and determine when the model can start failing. One of the possible constraints is using a different substrate for the graphene, as the substrate is known to affect the parameters on which the training of the model is based.[29] In our case the training dataset was obtained for graphene supported on $SiO_2$, so that the reliability of the model is expected to decrease for other substrates. As an example, Figure S13a-b shows the predictions for graphene supported on c-sapphire and on quartz, respectively. While for sapphire the results are still good, the performance for the quartz substrate decreased significantly, thus requiring of an additional training set for such substrate. The different processes to which graphene can be subjected, such as during the fabrication of devices, can also potentially alter its Raman spectrum. Fig. S13c shows the performance of the ML model on a BLG field-effect transistor, maintaining a reasonable accuracy. Another important factor can be the growth conditions of the graphene. Whereas our model was trained using graphene grown on thin Cu-Ni film substrates supported on sapphire,[18] the graphene of Fig. S13d was grown on a commercial Cu foil under different CVD conditions. These differences are evidenced in the optical images, with the BLG of Fig. S13d being present only at the center of isolated SLG grains, instead of the high coverage BLG used for the training set. The accuracy of the ML prediction decreases below a 95 %, with most of the errors occurring in the limit between areas with different stacking. It should be noted here that the specific classes present in this area (SLG, 20 – 28° and 28 – 30°) are those that the model is more prone to mix up (Fig. S8), making this a borderline case.

Finally, it is worth mentioning that this sensitivity of the Raman spectra towards the specific condition of the graphene can also be used as an opportunity to expand the capacities of the ML model. Thus, with appropriate training sets it seems conceivable to train models that can also determine the amount of strain or doping of the graphene,[30] or even the stacking order of graphene with respect to other 2D layered materials used as substrates.[47]

**Improvement of the precision of the twist angle determination.** Although the method proposed provides high accuracies for the determination of the twist angle of tBLG, the angle ranges of the different classes are still broad for most real applications. It is thus important to determine whether smaller ranges for the predicted angles can be potentially achieved. By using clustering analysis, we first tried to look for hidden information in our training set that can arise from different twist angles within each of the classes. Clustering comprises a range of unsupervised learning techniques that allow finding groups (clusters) of instances in the dataset without a prior knowledge of the classes.[40,41,48] Instances within each cluster are expected to be more similar to each other than to those in other clusters. We chose to use spectral clustering, an algorithm that permits to decide the number of clusters, hence allowing us to start from the same number of classes that we have determined. For simplicity, here the data of SLG and BLG-AB was removed from the dataset. Hence, the dataset now includes only the six classes of tBLG. Figure 7a shows the results after clustering the dataset using six clusters. These six clusters mainly correspond to the manual classification, with the only difference being a small mixture of the manual classes 20 – 28° and 28 – 30°. Except for this, each of the original classes is represented by a single cluster, as can be appreciated in Figure S14. Increasing the number of clusters resulted in some of the classes being subdivided into several clusters, as shown in Fig. 7b,c for seven and twelve clusters, respectively. Although it is not possible to determine the nature of these new clusters, the fact that the original classes do not share clusters seems to indicate that they might originate in differences in the twist angle within the given class. This opens an opportunity to increase the precision of the angle ranges of the supervised ML models by increasing the number of classes of the training dataset.

One way to attain this is to use a more precise determination of the twist angles of the training dataset, for example by TEM measurements.[9] This would allow to narrow some of the angle ranges of the training classes, hence supplying more information for the ML models. In this sense the R and R' bands seem promising candidates to increase the accuracy in the range of $5 - 20°$, as their shift and intensities are known to be affected by the twist angle.[9,49] Our data also points to this possibility, with the shift of the R band changing for different spectra as shown in Fig. S1. Correspondingly, some other overlooked trends in the features might be used by the models to increase the precision while keeping high accuracies. Another unexplored possibility to narrow down the predicted angle ranges is to use complementary excitation wavelengths for the Raman measurement. The trend of the characteristics of some Raman bands, such as $\omega$, *FWHM* and relative intensity, can largely depend on the wavelength of the excitation laser.[7,27] This provides additional information that can be used to obtain narrower angles for the classes of the training spectra, with accuracies that can be of the order of 0.1° for certain angle ranges.[27] Similarly, additional Raman bands and the inclusion of different regions of the spectra can also be used to increase the angle accuracy, especially for certain angle ranges.[25] Finally, another possibility to enhance the precision can be the use of whole (or parts) of the Raman spectra. This has already been proved with spectra simulated by first principles,[39] and more recently with trainsets limited to a few selected angles of artificially stacked tBLG.[40] However, the method employed here using only selected features of the spectra results in much faster and simpler calculations.

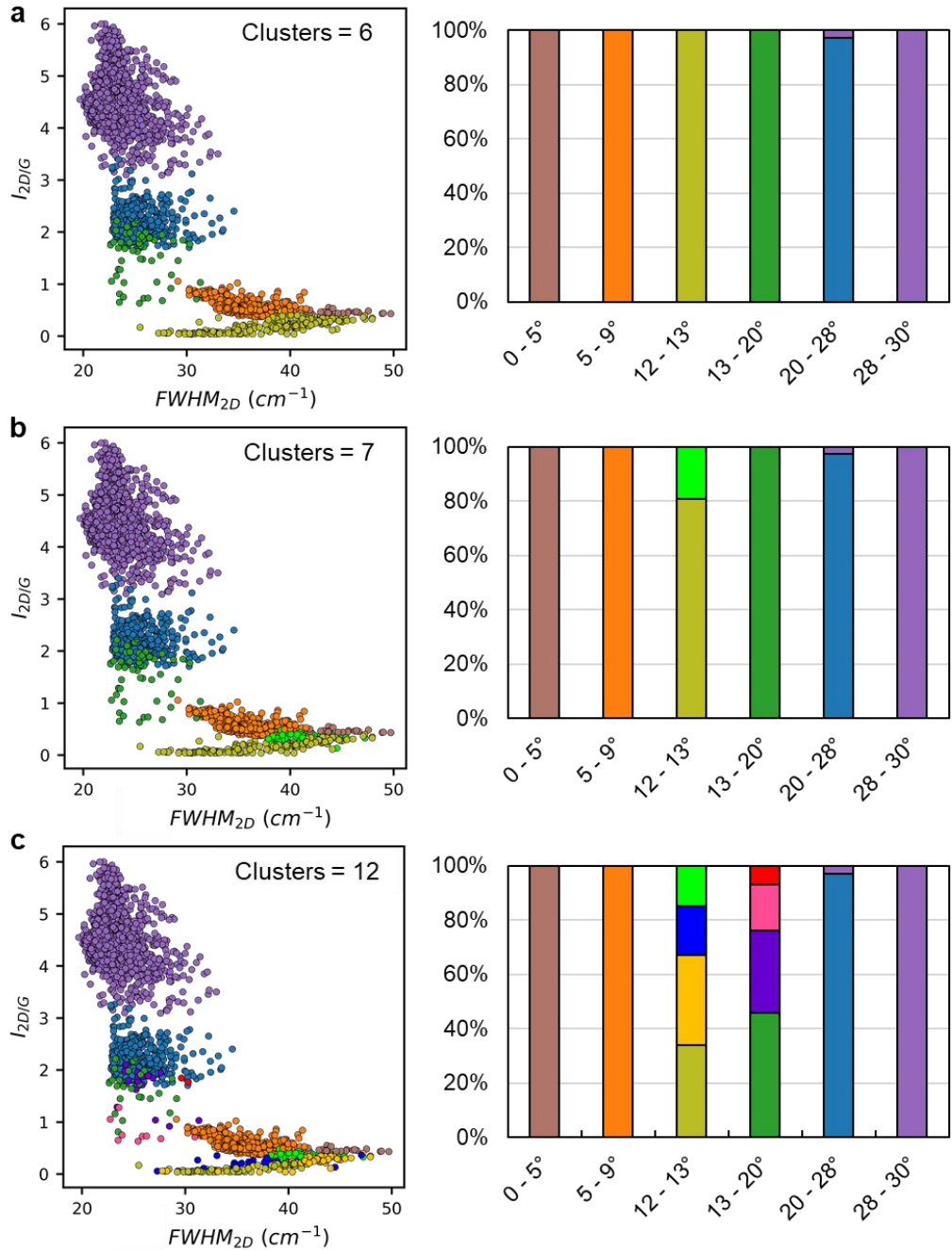

**Figure 7**. Clustering results of the dataset with only the tBLG classes for (a) 6, (b) 7 and (12) clusters. The left column shows the $FWHM_{2D}$ vs $I_{2D/G}$ scatter plot of the dataset, with each color corresponding to one of the found clusters. The right column shows how the original classes relate to the clusters found.

**Conclusions**

We have demonstrated the feasibility to integrate ML-based methods for the automated determination of the twist angle of CVD-grown tBLG. The proposed method provides accuracies that exceed a 99 % when compared with the manual labelling of the twist angles. Moreover, the method is not computationally demanding, providing predictions for whole Raman mappings comprising hundreds of spectra in a matter of seconds even on average desktop computers. The precision of the predicted angle ranges can be possibly increased by improving the quality of the data used for training the model. Finally, the simplicity of the proposed method make reasonable to expand it to determine the amount of strain and doping of graphene,[30] and the twist angle of other 2D materials[50] and even of heterostacks.[47] Paired with the flexibility and non-invasive nature of Raman measurements, we expect that this method facilitates the research of tBLG and of its intriguing properties and potential applications.

**Methods**

The BLG was grown by CVD on Cu-Ni thin films supported by sapphire crystals, at a temperature of 1085 °C.[18] An 80 μm thick Cu foil (Nilaco Co.) was used for the growth of isolated graphene grains at 1050 °C. After the CVD, graphene was transferred to an $SiO_2$ substrate using a PMMA support film. Confocal Raman spectroscopy was conducted on the transferred graphene using a 532 nm laser excitation in a Nanofinder 30 spectrometer (Tokyo Instruments Inc.).

The complete training dataset was generated from ~6000 individual Raman spectra collected from graphene samples obtained from 8 different CVD batches. The open-source Python library scikit-learn was used to perform the ML modeling.[42] The dataset was used without any processing except for the SVM, multilayer perceptron (neural network) and k-nearest neighbors models, for which each feature was centered around the mean and then scaled to unit variance. To optimize the hyperparameters of each of the models the original dataset was split into a training and a test subsets. The new training subset was then used to find the best hyperparameters via stratified k-fold cross validation,[45] while the held-out test subset was only used to evaluate the performance of the different models.[51] The accuracy of the trained models was finally obtained by comparing predictions of the model and the manual labelling of a different set of Raman mappings not used during the training process.

Before the unsupervised clustering, the data for SLG and BLG-AB was removed from the training set, and each feature was scaled by its maximum absolute value. A neighborhood components analysis transformation with 2 components was then performed on the dataset, to differentiate the manually determine classes.[44] The clustering algorithm employed was a spectral clustering, which allows to decide the number of clusters beforehand.[48]

**Corresponding Authors**

*Email: pablosolisfernandez@gmail.com

*Email: h-ago@gic.kyushu-u.ac.jp

**Acknowledgements**

This work was supported by the JSPS KAKENHI grant numbers JP18H03864, JP19K22113, 21K18878, and JSPS Transformative Research Areas (A) "Science of 2.5 Dimensional Materials" program (21H05232, 21H05233), JST CREST grant numbers JPMJCR18I1, JPMJCR20B1 and the JSPS A3 Foresight Program.

# Supplementary Information

# Determining the Twist Angle of Bilayer Graphene by Machine Learning Analysis of its Raman Spectrum

*Pablo Solís-Fernández, \*,† Hiroki Ago\*,†,‡*

† Global Innovation Center (GIC), Kyushu University, Fukuoka 816-8580, Japan

‡ Interdisciplinary Graduate School of Engineering Science, Kyushu University, Fukuoka 816-8580, Japan

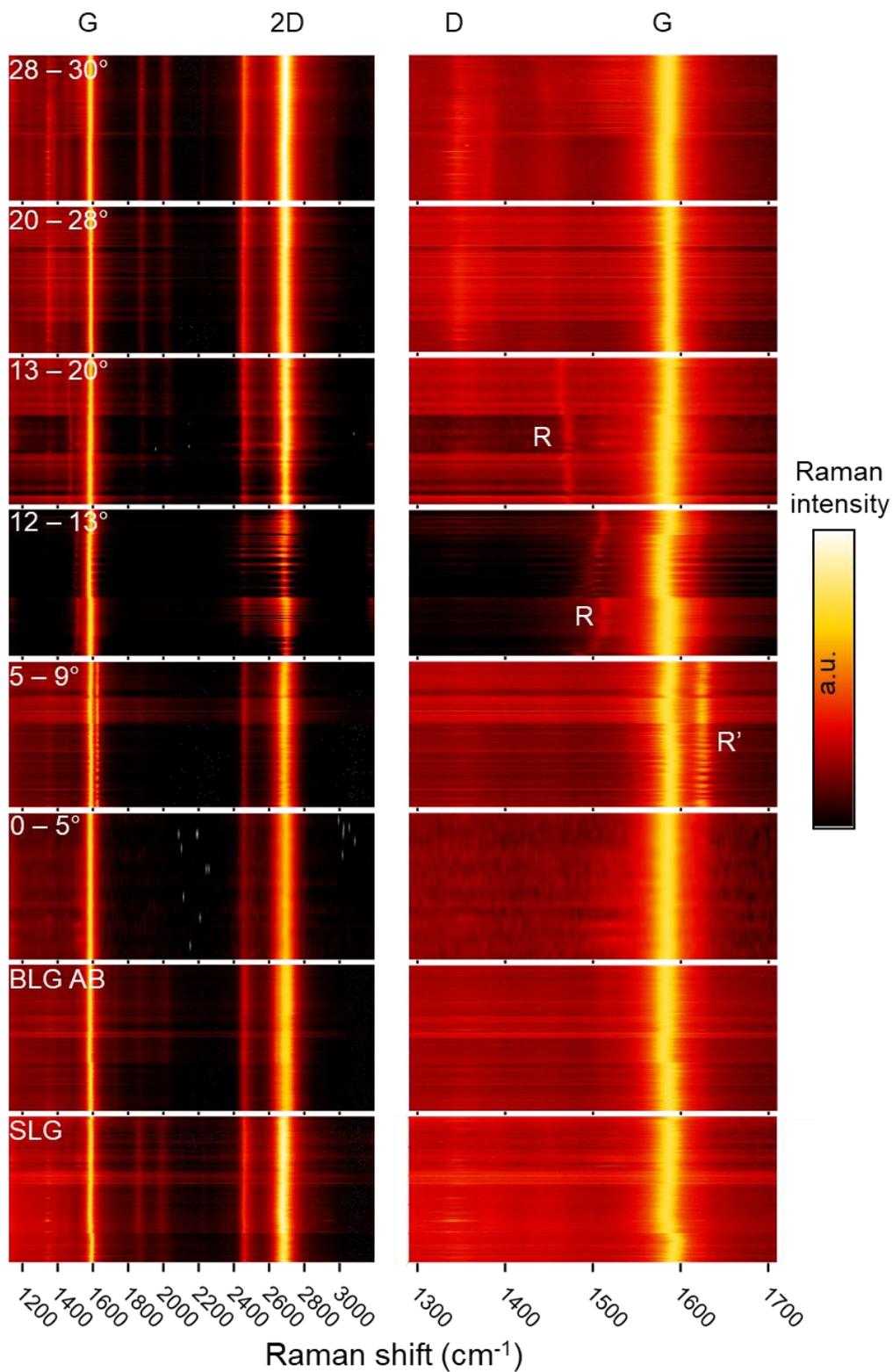

**Figure S1**. Plots of all the Raman spectra used to extract the features used for the manual classification and for the training of the ML models. Each individual spectrum corresponds to a horizontal line, with the color representing the Raman intensity. The left column corresponds to the whole spectra, while the right column are enlargements around the G-band area. The spectra are normalized to the G band intensity, and the color scale is logarithmic to better visualize the less intense bands. The most relevant bands are labelled.

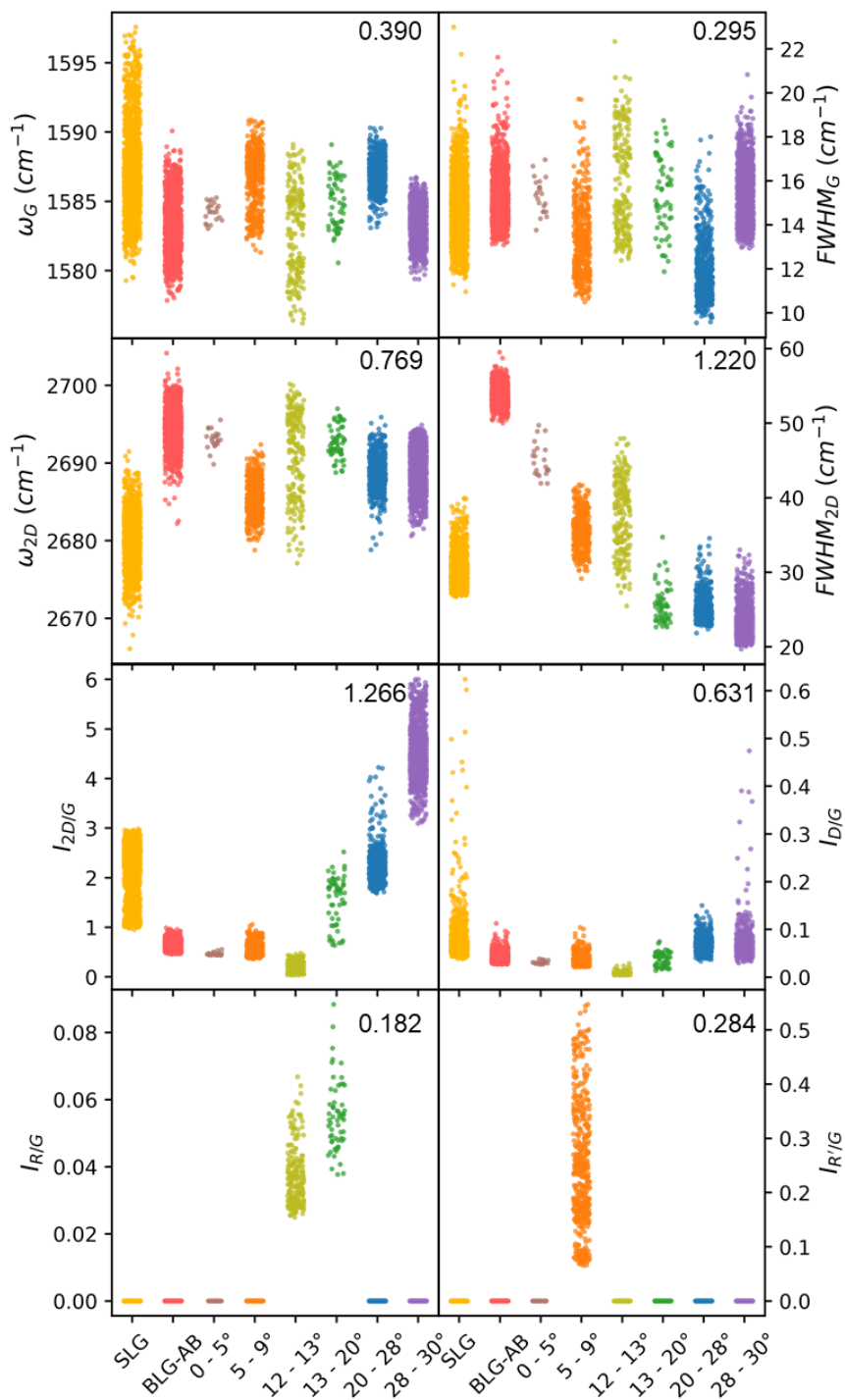

**Figure S2**. Distribution of all the features used to train the ML models. This data was experimentally collected from spectra taken from several CVD-grown graphene samples transferred to $SiO_2$ substrates. The numbers at the top right corners are the values for the mutual information between each feature and the stacking angle classes.

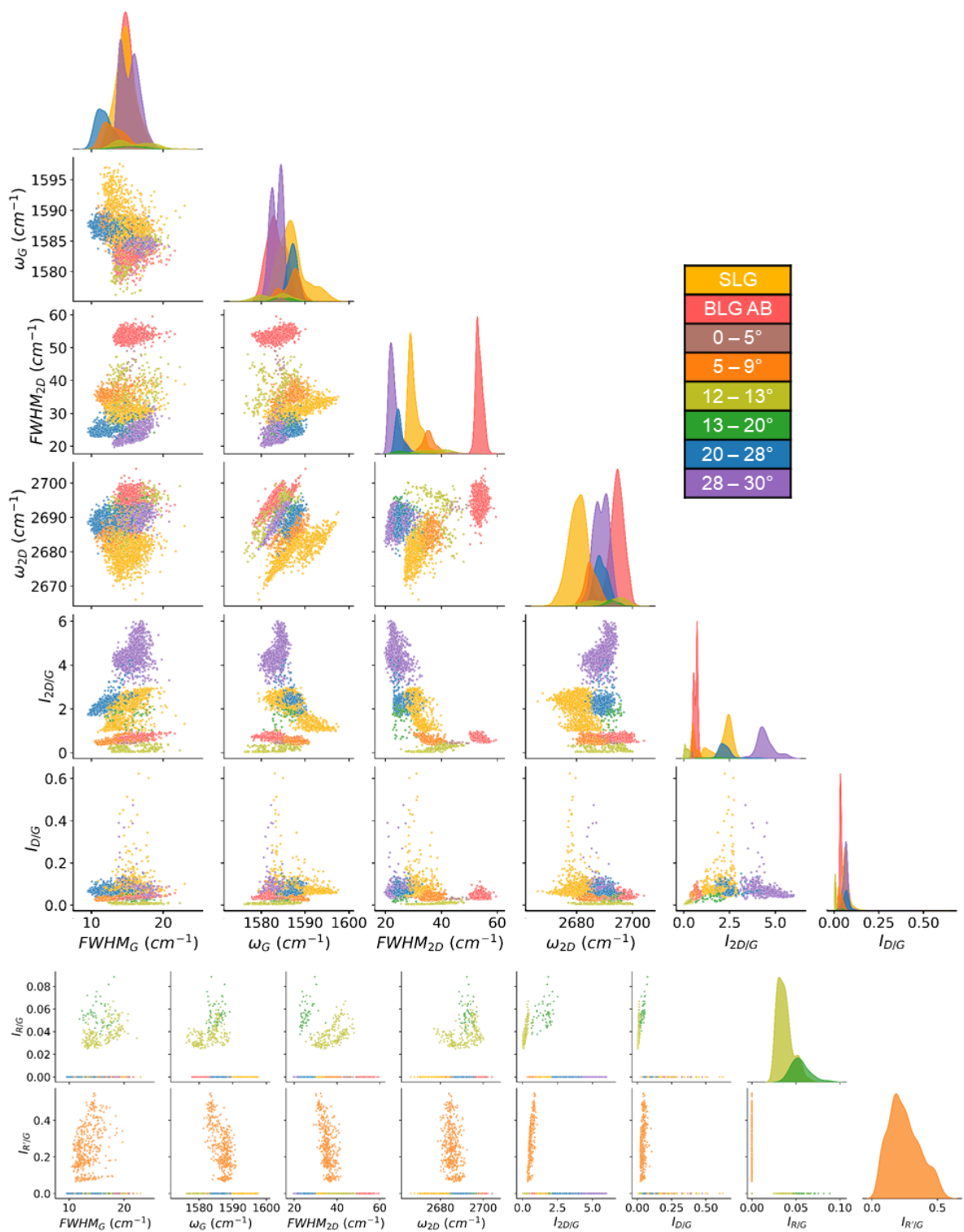

**Figure S3.** Scatter plots for each pair of features, with the probability distribution of each feature being shown in the top diagonal. The data is taken from the whole set of spectra used as training set. The last two rows have been separated from the rest for the sake of clarity.

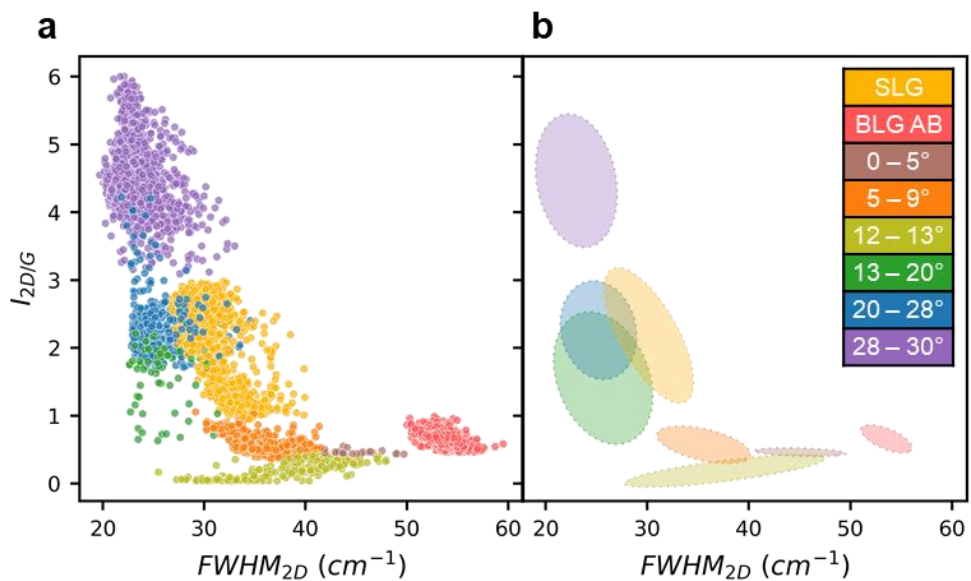

**Figure S4**. (a) Scatter plot of the $FWHM_{2D}$ and the $I_{2D/G}$ for each of the spectrum in the dataset. (b) Areas enclosing ~95 % of the points for each of the stacking angle classes in (a). The partial overlapping of the ellipses in (b) shows that the unambiguous determination of the class is not possible when using only a pair of features.

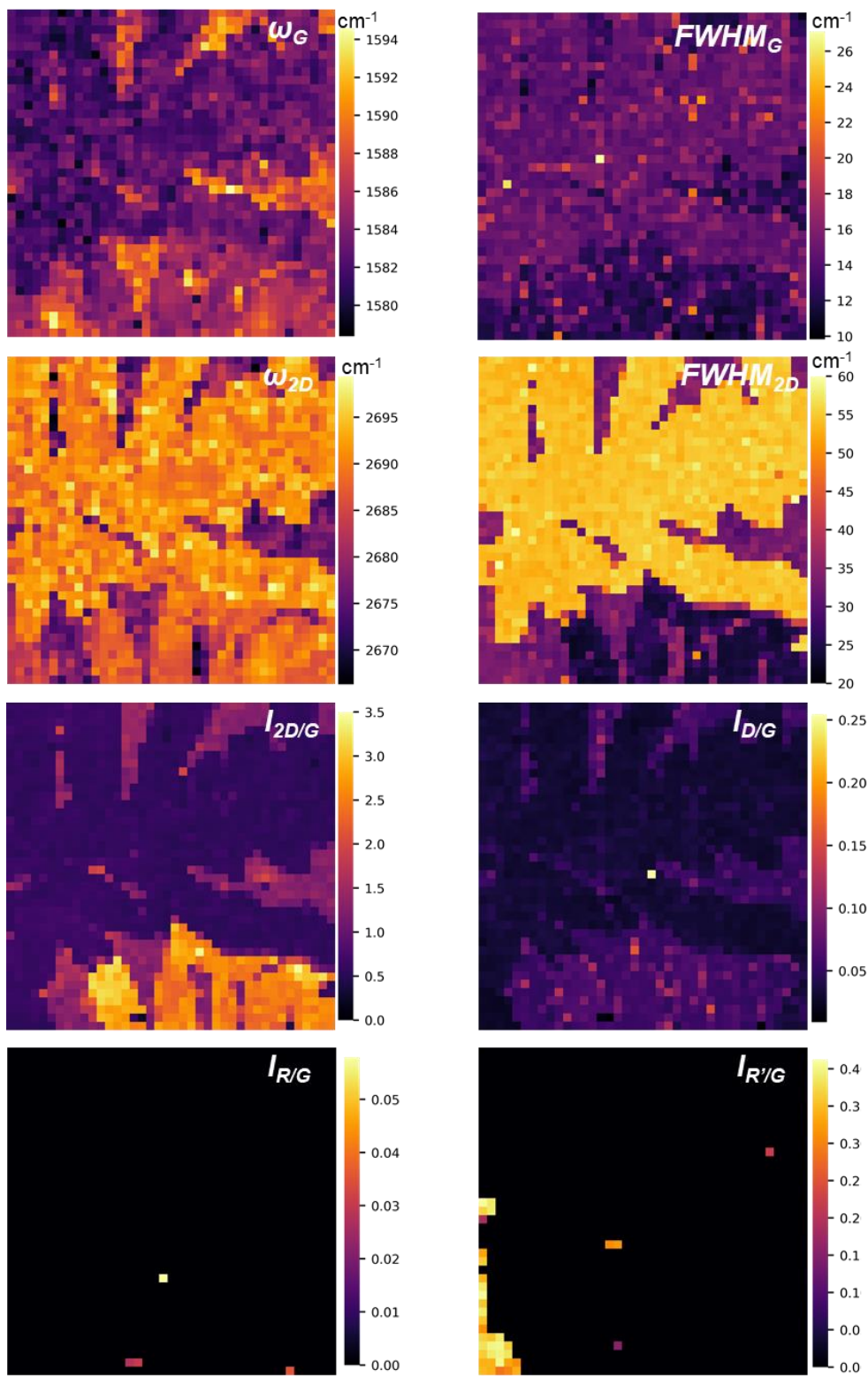

**Figure S5**. Raman mappings of all the features used to classify the Raman spectra of the graphene area shown in Figure 4a of the main text.

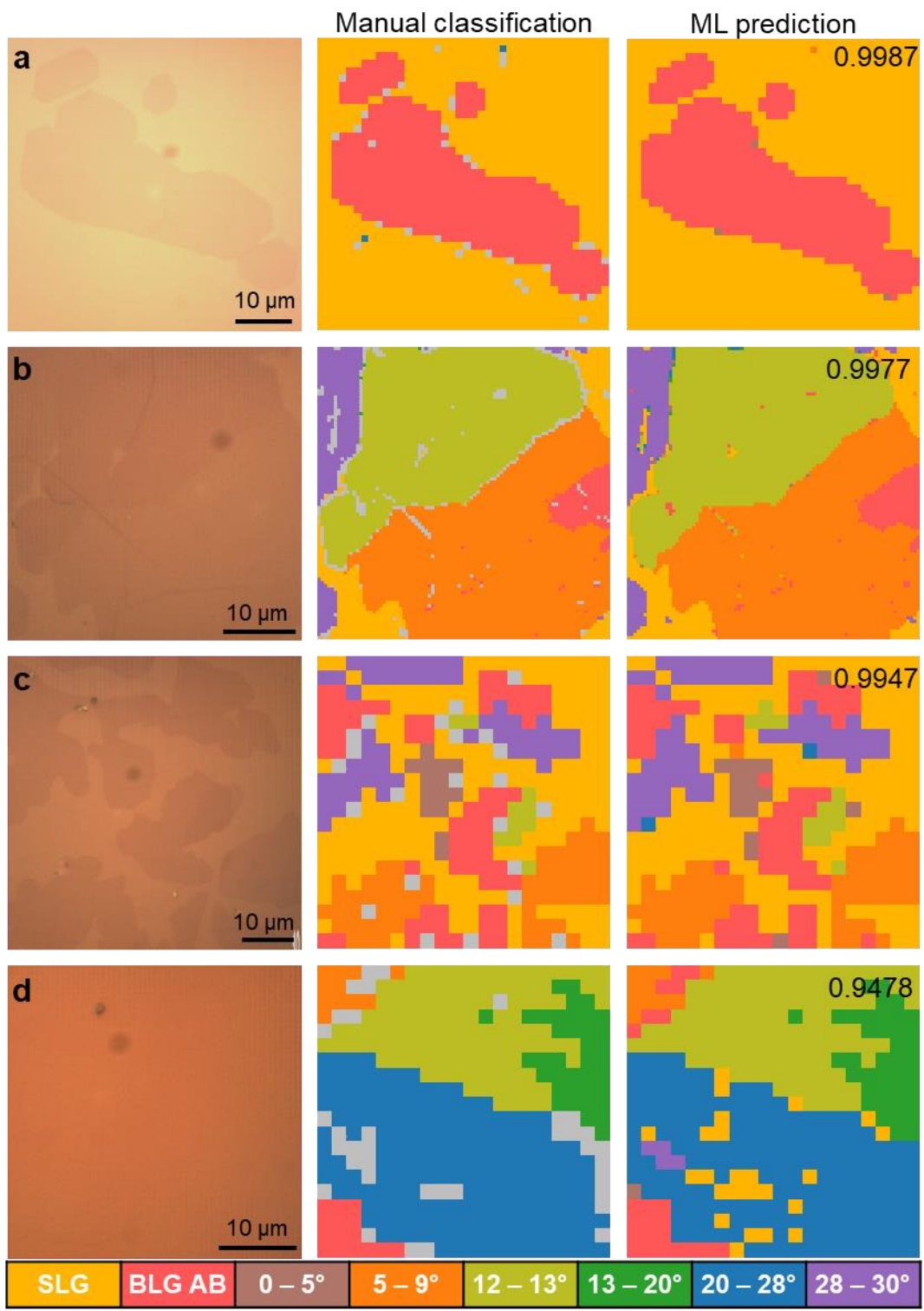

**Figure S6**. Results of the predictions of the random forest model for several areas. Each area was taken from a different sample. The left, center and right columns show the optical image, manual labelling, and the prediction of the ML model, respectively. The numbers of the top right corner represent the accuracy of the prediction based on the manual labelling.

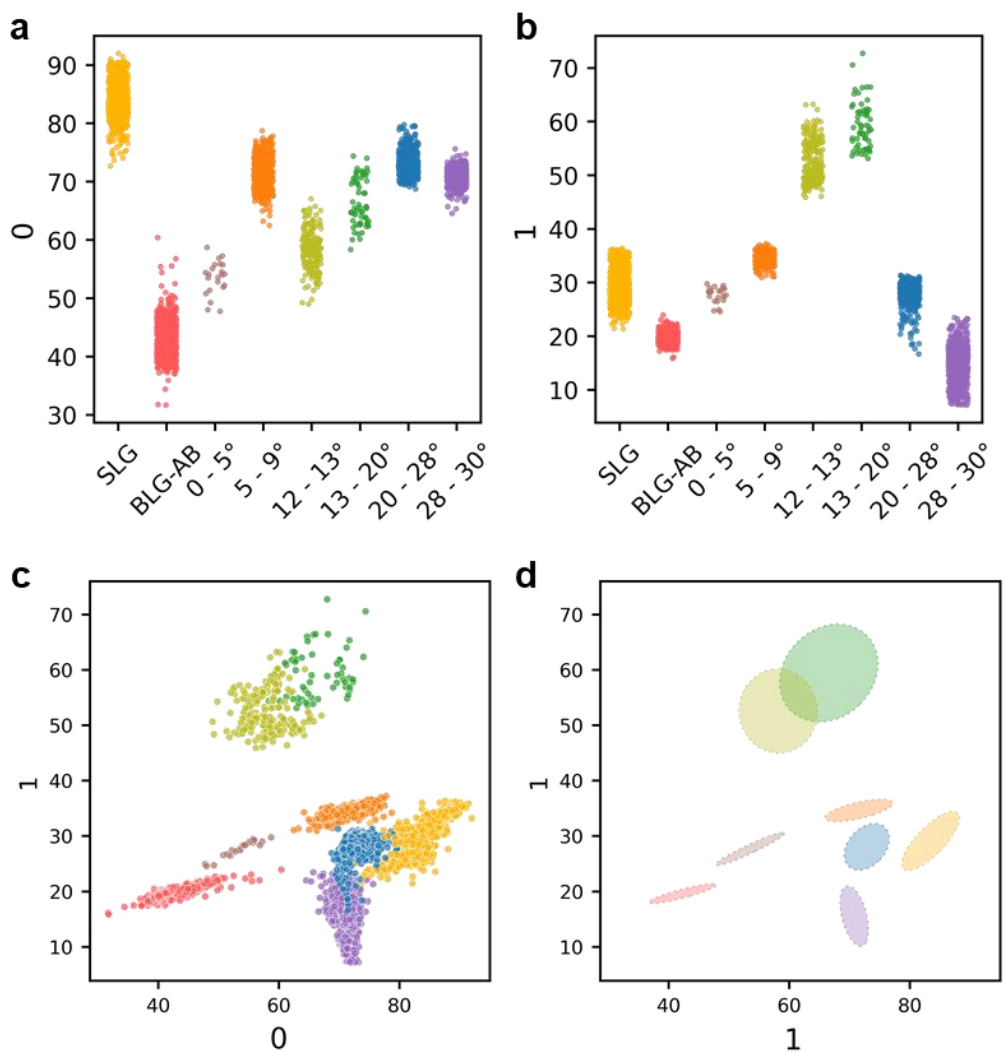

**Figure S7**. The dimensionality of the original dataset has been reduced from 8 to 2 features by applying a neighborhood components analysis algorithm. The new features 0 (a) and 1 (b) retain information from the original 8 features. Both features are plotted together in (c), with the ellipses in (d) showing the areas enclosing ~95 % of the points for each of the stacking angle classes. The overlapping in (d) is lower than that observed when using 2 of the original features (Fig. S4).

|  | ML prediction | | | | | | | | Sensitivity $T_P/(T_P+F_N)$ |
| --- | --- | --- | --- | --- | --- | --- | --- | --- | --- |
|  | SLG | BLG-AB | 0 - 5° | 5 - 9° | 12 - 13° | 13 - 20° | 20 - 28° | 28 - 30° |  |
| SLG | 0.9917 | 0 | 0 | 0.0029 | 0 | 0 | 0.0019 | 0.0034 | 0.992 |
| BLG-AB | 0 | 1 | 0 | 0 | 0 | 0 | 0 | 0 | 1.000 |
| 0 - 5° | 0 | 0 | 0.9333 | 0.0667 | 0 | 0 | 0 | 0 | 0.933 |
| 5 - 9° | 0.0003 | 0.0003 | 0 | 0.9994 | 0 | 0 | 0 | 0 | 0.999 |
| 12 - 13° | 0 | 0 | 0 | 0 | 0.9997 | 0.0003 | 0 | 0 | 1.000 |
| 13 - 20° | 0 | 0 | 0 | 0 | 0.0233 | 0.9767 | 0 | 0 | 0.977 |
| 20 - 28° | 0.0612 | 0 | 0 | 0.0026 | 0 | 0 | 0.926 | 0.0102 | 0.926 |
| 28 - 30° | 0.0016 | 0 | 0 | 0 | 0 | 0 | 0.0031 | 0.9953 | 0.995 |
| Specificity $T_N/(T_N+F_P)$ | 0.997 | 1.000 | 1.000 | 0.999 | 1.000 | 1.000 | 0.999 | 0.999 |  |
| Undetermined | 0.343 | 0.123 | 0.048 | 0.133 | 0.231 | 0.010 | 0.064 | 0.048 |  |

(Manual classification on rows)

**Figure S8**. Confusion matrix for the ML predictions in Fig. 4e of the main text and Fig. S6. The values are normalized by rows. The sensitivity (or true positive rate) and specificity (or true negative rate) of each class are also shown, where $T_P$ and $F_P$ ($T_N$ and $F_N$) are the number of true and false positives (negatives), respectively. The last row represents the ratio of classes predicted by the ML model for the spectra that could not be manually classified.

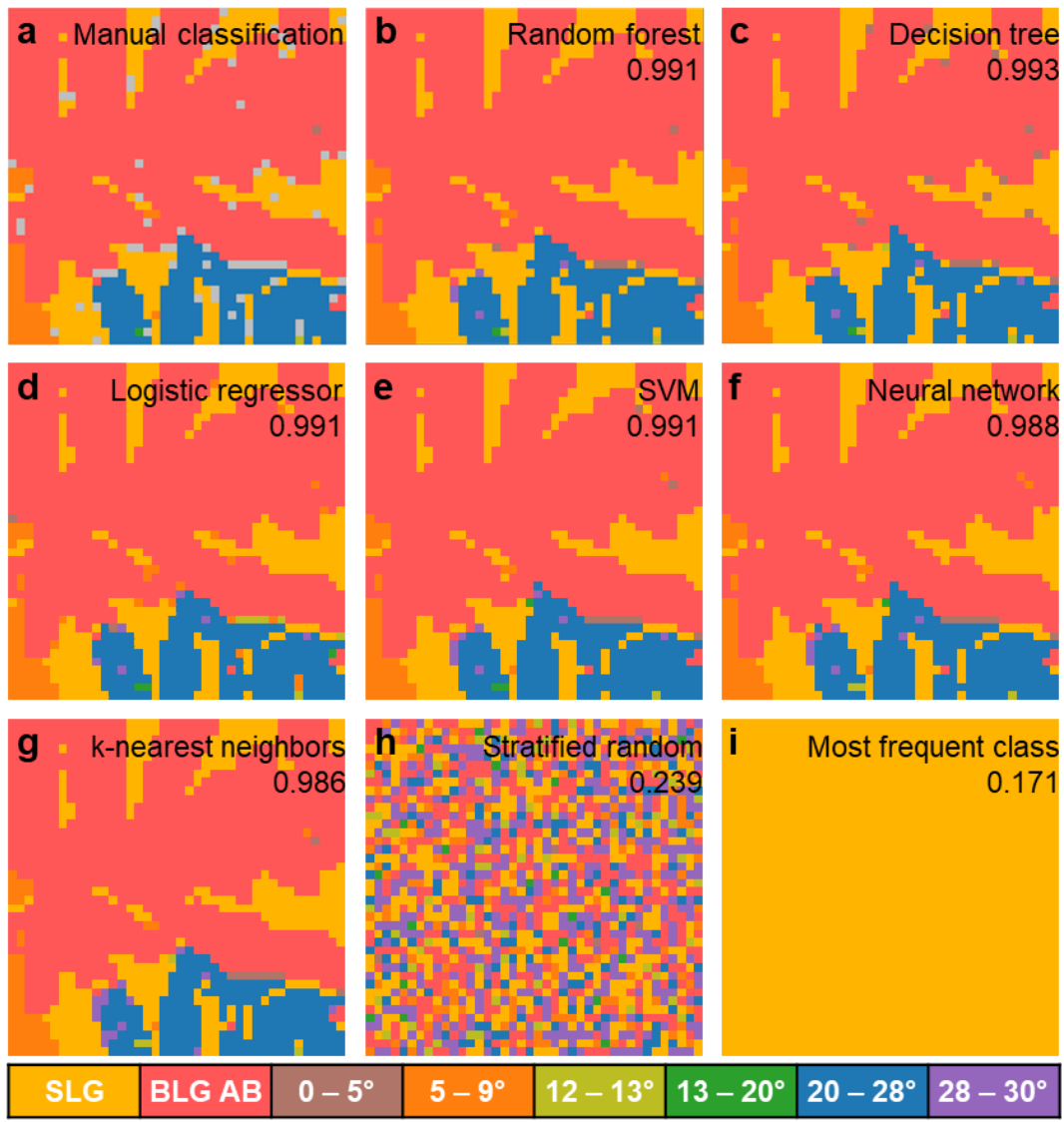

**Figure S9**. Results of the prediction of different ML models for the same mapping data used in the figure 4 of the main text. The top right corner of each image shows the specific classifier used, and the number in parenthesis is the accuracy when comparing the results with the manual classification in (a). The accuracy values exclude the manually undetermined Raman spectra (gray points in (a)). A couple of dummy models have also been included ((h) and (i)) to test the efficiency of the real models. For these dummy models, the predictions are independent of the training data. The predictions of (h) are random but respect the class distribution of the train dataset, while in (i) the most frequent class in the train dataset is always predicted.

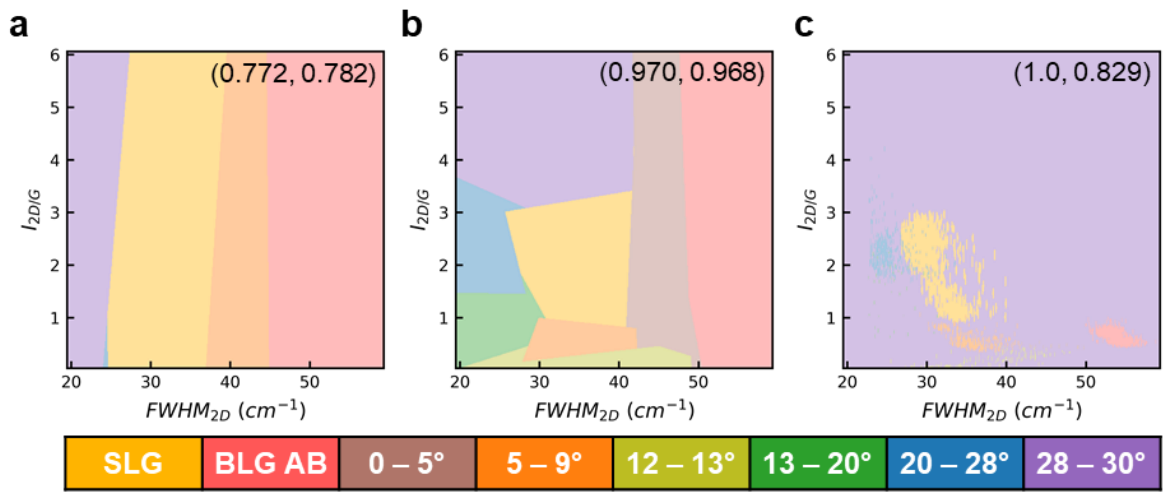

**Figure S10**. Decision boundaries for underfitted (a), well-fitted (b) and overfitted (c) SVM models. All the models were trained using the two most relevant features ($I_{2D/G}$ and $FWHM_{2D}$). The three different behaviors are attained by properly adjusting the hyperparameters. The numbers in parenthesis at the top right corners represent the accuracy of the model on the train and test sets, respectively. The decision boundaries of the under and overfitted models look very different to those of a properly trained model. The underfitted model (a) cannot identify all the different classes in the train set, having similarly low accuracy values for the train and test sets. The overfitted model (c) has complex boundaries, and the accuracy of the test set significantly drops respect to that of the trainset. This indicates the poor ability of the overfitted model to generalize to unseen data. The decision boundary for the well-fitted SVM is qualitatively similar to that of the well-fitted random forest of Fig. 5 of the main text. For a better comparison, the training features of the SVM models here were not scaled.

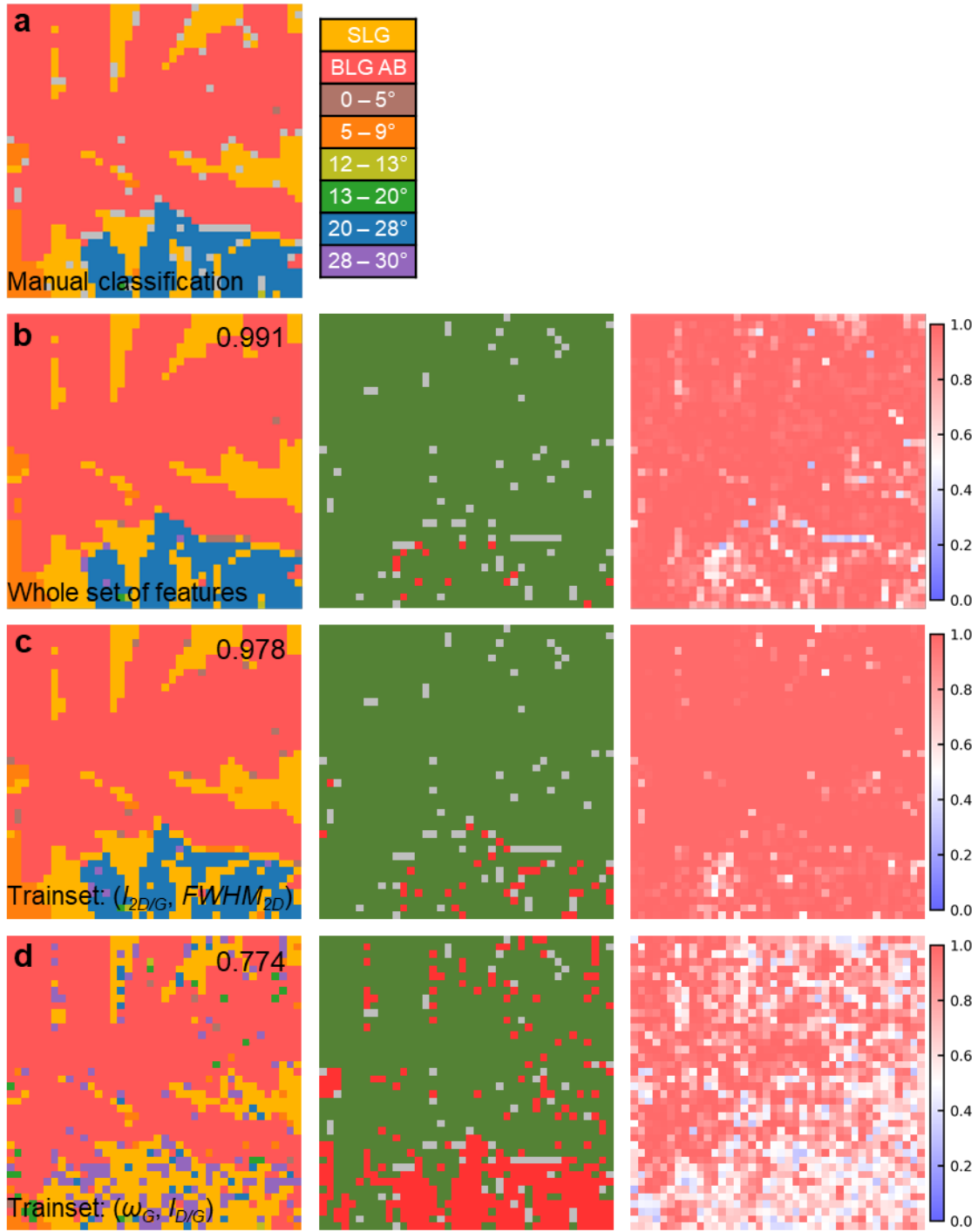

**Figure S11.** Comparison of the manual classification (a), with the prediction results when training the ML model using all the features (b), the two most relevant features ($I_{2D/G}$ and $FWHM_{2D}$) (c), and two non-ideal features ($\omega_G$ and $I_{D/G}$) (d). The first column represents the predictions of the corresponding models, the middle column show the differences compared with the manual classification (in red), and the last column are the confidence scores of each model. The numbers in the top right corner of the first column are the corresponding accuracies.

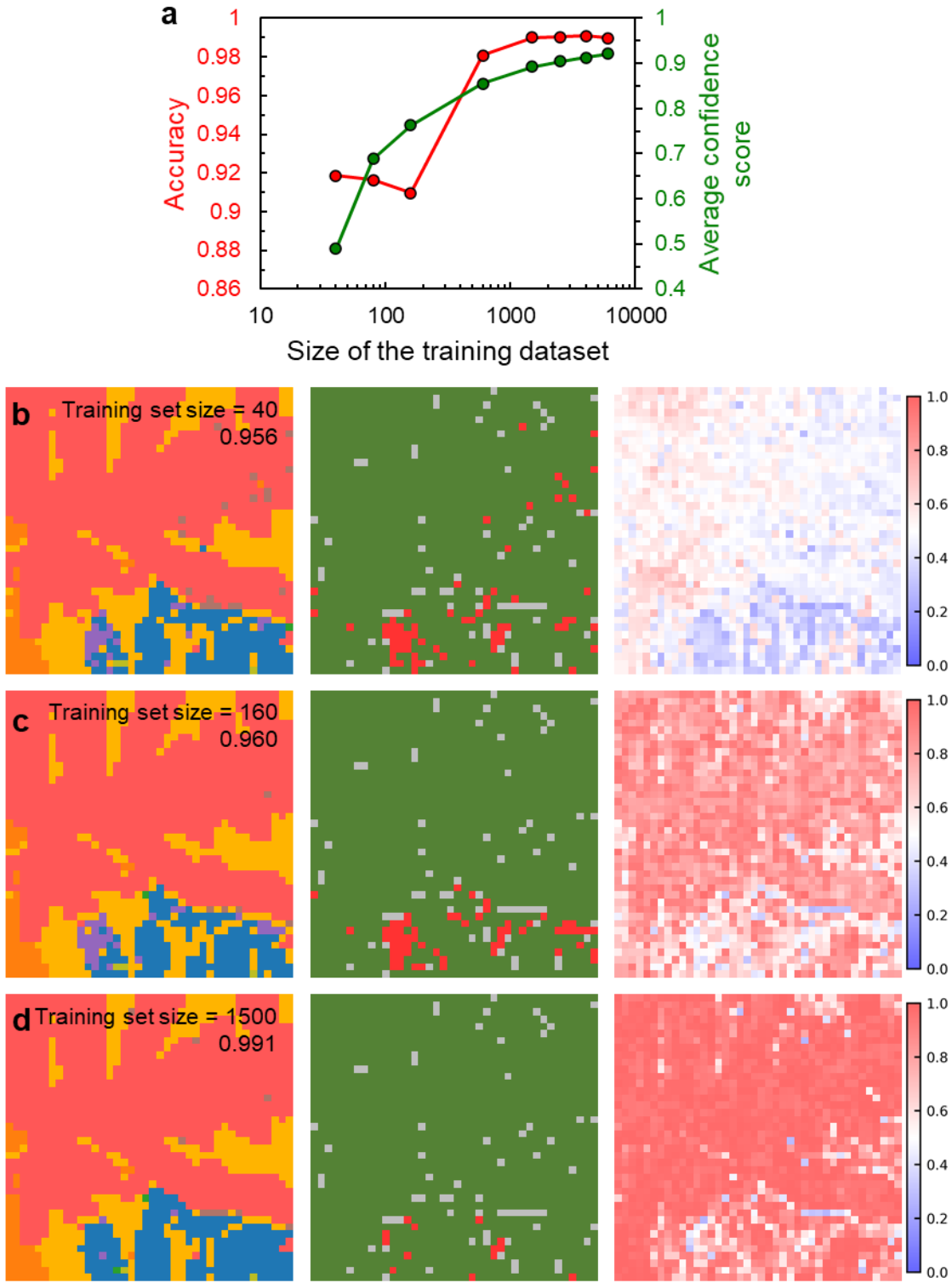

**Figure S12**. (a) Average accuracy and probability of the assigned class as a function of the training set size. Smaller sets were extracted from the original one and used to evaluate the areas in Fig. 4 and S12. (b-d) Results of the prediction of datasets with different sizes for the area of Fig. 4. The first column in (b-d) represents the predictions of the corresponding models, with the top right corners showing the size of the set and the accuracy of the prediction for this region. The middle and last column in (b-d) show the differences compared with the manual classification (in red), and the confidence scores of each model, respectively.

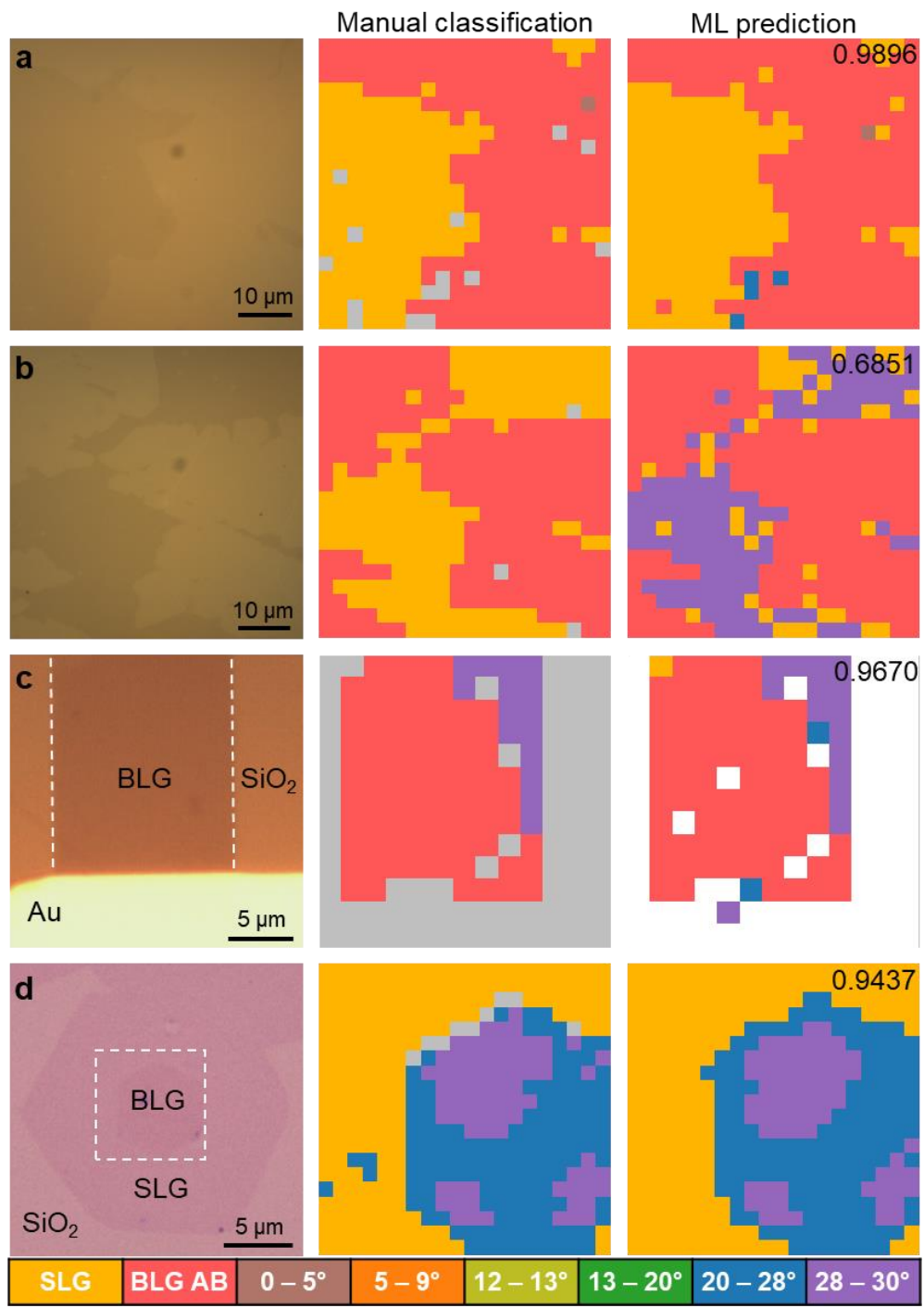

**Figure S13**. Results of the predictions from the random forest model for BLG under different conditions. (a, b) BLG on c-sapphire (a) and quartz (b) substrates. (c) Channel of a BLG FET device (delimited by the dashed line). (d) Isolated SLG grain with a BLG island on the center, grown on using a Cu foil. The dashed square in (d) shows the area mapped by Raman.

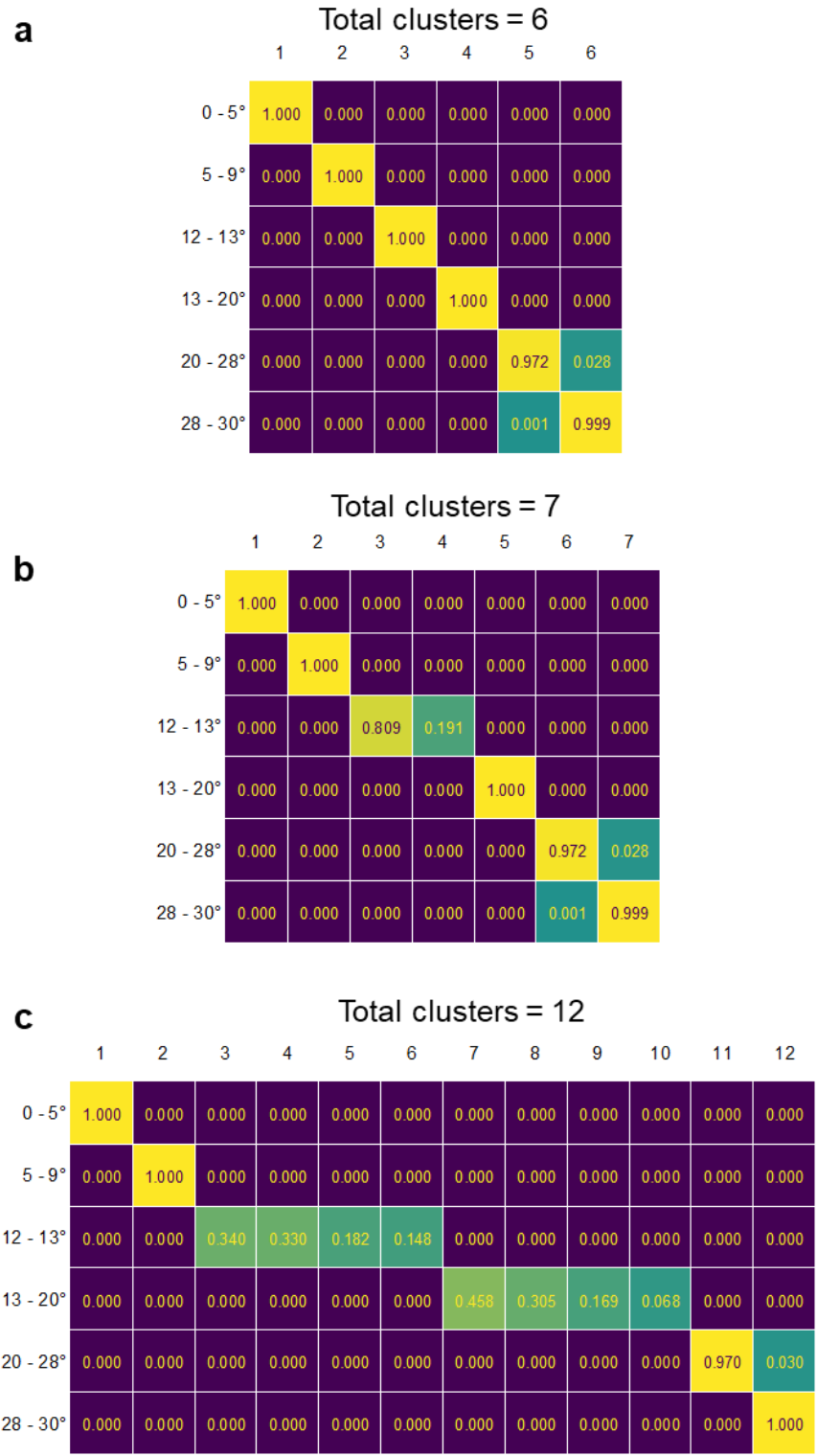

**Figure S14**. Relation between the original classes of the dataset (rows) and the different clusters found in Fig. 7 of the main text (columns). The total number of clusters are (a) 6, (b) 7 and (c) 12. The numbers are normalized by class (row), representing the ratio of instances of a class belonging each cluster. Except for the small mix between the classes 20 – 28° and 28 – 30°, each of the classes in (a) is represented by a single cluster. Increasing the total number of clusters results in some of the classes being subdivided in clusters.